\documentclass[proof]{pasj00}
\begin{document}
\SetRunningHead{K. Yoshikawa et al.}{Warm/Hot Intergalactic Medium}
\title{Locating the Warm-Hot Intergalactic Medium \\
in the Simulated Local Universe}

\author{
Kohji \textsc{Yoshikawa}\altaffilmark{1},
Klaus \textsc{Dolag}\altaffilmark{2},
Yasushi \textsc{Suto}\altaffilmark{1,3},
Shin \textsc{Sasaki}\altaffilmark{4},
Noriko Y. \textsc{Yamasaki}\altaffilmark{5},\\
Takaya \textsc{Ohashi}\altaffilmark{4},
Kazuhisa \textsc{Mitsuda}\altaffilmark{5},
Yuzuru \textsc{Tawara}\altaffilmark{6},
Ryuichi \textsc{Fujimoto}\altaffilmark{5},\\
Tae \textsc{Furusho}\altaffilmark{5},
Akihiro \textsc{Furuzawa}\altaffilmark{6},
Manabu \textsc{Ishida}\altaffilmark{4},
Yoshitaka \textsc{Ishisaki}\altaffilmark{3}, 
and
Yoh \textsc{Takei}\altaffilmark{5}
}

\altaffiltext{1}{Department of Physics, School of Science, 
The University of
Tokyo, Tokyo 113-0033}
\email{kohji@utap.phys.s.u-tokyo.ac.jp} 

\altaffiltext{2}{Dipartimento di Astronomia, Universit\`a di Padova, 
\\ vicolo dell'Osservatorio 5, 35122 Padova, Italy}

\altaffiltext{3}{Research Center for the Early Universe (RESCEU), School
of Science, \\ The University of Tokyo, Tokyo 113-0033}

\altaffiltext{4}{Department of Physics, Tokyo Metropolitan University,
\\ 1-1 Minami-Osawa, Hachioji, Tokyo 192-0397}

\altaffiltext{5}{The Institute of Space and Astronautical Science
(ISAS), \\ Japan Aerospace Exploration Agency (JAXA), \\
3-1-1 Yoshinodai, Sagamihara, Kanagawa 229-8510}

\altaffiltext{6}{Department of Physics, Nagoya University, 
Furo-cho, Chikusa-ku, Nagoya 464-8602}
\Received{2004/07/14}
\Accepted{2004/??/??}

\maketitle


\begin{abstract}
 We present an analysis of mock spectral observation of warm-hot
 intergalactic medium (WHIM) using a constrained simulation of the local
 universe. The simulated map of oxygen emission lines from local WHIM
 reproduces well the observed structures traced by galaxies in the real
 local universe. We further attempt to perform mock observations of
 outer parts of simulated Coma cluster and A3627 adopting the expected
 performance of {\it DIOS} (Diffuse Intergalactic Oxygen Surveyor),
 which is proposed as a dedicated soft X-ray mission to search for
 cosmic missing baryons.  We find that WHIMs surrounding nearby clusters
 are detectable with a typical exposure time of a day, and thus
 constitute realistic and promising targets for {\it DIOS}.  We also
 find that an X-ray emitting clump in front of Coma cluster, recently
 reported in the XMM-Newton observation, has a counterpart in the
 simulated local universe, and its observed spectrum can be well
 reproduced in the simulated local universe if the gas temperature is
 set to the observationally estimated value.
\end{abstract}

\section{Introduction}

It is widely accepted that our universe is dominated by {\it dark}
components; 23 percent in dark matter, and 73 percent in dark energy
\citep{Spergel03}.  More surprisingly, even the remaining 4 percent,
cosmic baryons, has largely evaded the direct detection so far, i.e.,
most of the baryons is indeed {\it dark}. While those {\it cosmic
missing baryons} may consist of compact stellar objects (white dwarfs,
neutron stars and black holes), brown dwarfs, and/or diffuse gas, recent
numerical simulations indicate that they largely take a form of the
warm-hot intergalactic medium (WHIM) with temperature of $10^5 {\rm K}
<T<10^7 {\rm K}$ (e.g., \cite{Cen1999a}). In fact cosmological
hydrodynamic simulations consistently point to the fact that WHIM traces
the large-scale filamentary structure of dark matter distribution more
faithfully than hot intracluster gas ($T > 10^7 {\rm K}$) and galaxies
both of which preferentially resides in clusters that form around the
knot-like intersections of the filamentary regions. This implies that
WHIM carries important cosmological information in a complementary
fashion to distribution of galaxies (in optical) and of clusters (in
X-ray).

Unfortunately the X-ray emission of WHIM via the thermal bremsstrahlung
is very weak, and its detection has been proposed only either through
O{\sc vii} and O{\sc viii} absorption features in the QSO spectra or the
possible contribution to the cosmic X-ray background in the soft
band. While there are several observational reports to have detected
signatures of WHIM via oxygen absorption lines \citep{Nicastro2002,
Fang2002, Mathur2003} and via excess X-ray emissions \citep{Kaastra2003,
Finoguenov2003} among others, they are not well suited for unbiased
exploration of WHIM that is important for cosmological studies.  In
order to identify the existence, and explore the nature, of elusive
cosmic missing baryons in a more unambiguous manner, we propose a
dedicated soft X-ray mission, {\it DIOS} (Diffuse Intergalactic Oxygen
Surveyor; see Ohashi et al. 2004) which will carry out a direct and
homogeneous survey of WHIM. In a previous paper (Yoshikawa et al. 2003;
Paper I hereafter), we examined in detail the detectability of WHIM on
the basis of the cosmological hydrodynamic simulations. We indeed showed
that the unprecedented energy resolution ($\sim 2$eV) of the X-ray
Spectrometer Array (XSA) on-board {\it DIOS} enables us to identify WHIM
with gas temperature $T=10^{6-7}$K and overdensity $\delta=10-100$
located at redshift $z<0.3$ through emission lines of O{\sc vii} and
O{\sc viii}.

The purpose of the present paper is to consider the detectability of
WHIM surrounding nearby clusters ($z<0.03$) using the constrained
numerical simulation to reproduce the local universe \citep{Dolag2003}.
This approach is useful in setting the observational strategy of the
targeted observations around the previously known structures.  This is a
complementary and equally important observing mode to the blind survey
that is implicitly kept in mind in Paper I. 

It should be noted here that after submitting the first manuscript of
the present paper, we were informed of an earlier paper discussing
exactly the same topic by \citet{Kravtsov2002}. They used the
constrained simulation of the nearby universe based on
\citep{Klypin2003} the MARK III survey of peculiar velocities, and
addressed the observational signatures including emission by O{\sc vii}
and O{\sc viii}. Since they do not specifically bear in mind the
high-resolution spectroscopy which will be feasible with {\it DIOS},
they do not study the detailed emission spectra that we will present
below. Nevertheless the overall qualitative features that they found are
in good agreement with our results.

The rest of the paper is organized as follows; section 2 describes the
local universe simulation data that we use for mock observations.  In \S
3, we summarize how to compute the soft X-ray spectrum of the local
universe with the expected performances of {\it DIOS} following Paper I.
Several examples of mock WHIM observation are shown in \S 4. We focus on
the mock observation around the outskirts of the simulated Coma cluster
in \S 5, and show that our simulated spectrum reasonably reproduces a
recent observational report of the XMM-Newton discovery of an X-ray
clump in Coma by \citet{Finoguenov2003}. Finally, section 6 is devoted
to the conclusions of the paper.

\section{Simulation of the Local Universe}

Throughout this paper we analyze a dataset generated by a cosmological
hydrodynamic simulation which attempted to reproduce our local
universe. The simulation is based on the previous work by
\citet{Mathis2002} which intentionally constructed the initial
conditions making use of the IRAS 1.2-Jy survey \citep{Fisher1994,
Fisher1995} out to $cz<12000{\rm km~ s}^{-1}$. They assumed a
$\Lambda$CDM ($\Lambda$-dominated cold dark matter) model with the
density parameter $\Omega_{\rm m}=0.3$, the dimensionless cosmological
constant $\Omega_\Lambda=0.7$, the Hubble constant $h=0.7$ (in units of
100 km/s/Mpc), and the rms density fluctuation $\sigma_8=0.9$ (top-hat
smoothed over a scale of $8h^{-1}\mbox{Mpc}$). First the observed galaxy
density field was Gaussian smoothed on a scale of $5h^{-1}\mbox{Mpc}$.
Then it was evolved (quasi-)linearly back to the initial redshift of the
simulation, $z=50$. Its overall amplitude is normalized so that the
resulting rms mass variance matches the adopted value of
$\sigma_8$. This was used as a Gaussian constraint for an otherwise
random realization of the $\Lambda$CDM cosmology. Centered on the Milky
Way the size of simulation box is $240h^{-1}$Mpc, with a high resolution
region covering a sphere with $80h^{-1}$Mpc of radius which corresponds
to the region constrained by the observations. Evolved forward in time,
\citet{Mathis2002} demonstrated that this simulation very well
reproduces the local universe today in several respects; applying
semi-analytic galaxy formation scheme to their dark matter only
simulation very well reproduces the local galaxy population.  Density
and velocity fields obtained from synthetic mock galaxy catalogues
reproduce very well the observed counterparts. Also some of the most
prominent halos in the simulations may be identified and indeed agree
well in positions and masses with the observed galaxy clusters. Further
details of the construction of the initial condition are found in
\citet{Mathis2002}; see also independent earlier work by
\citet{Klypin2003} and \citet{Kravtsov2002}.

\citet{Dolag2003} repeated the  simulation including gas physics with
smoothed particle hydrodynamics (SPH) technique. They extended these
initial conditions by splitting the original high resolution dark matter
particles into gas and dark matter particles with masses of $0.48 \times
10^9h^{-1}M_\odot$ and $3.1 \times 10^9h^{-1}M_\odot$, respectively. The
gravitational force resolution of the simulation was set to be
$7.5h^{-1}{\rm kpc}$ which is also the resulting mean inter-particle
separation of the SPH particles in the dense centers of the simulated
galaxy cluster. The most massive cluster in this simulation is thereby
resolved by nearly one million particles.

This simulation was carried out with {\small GADGET-2}, a new version of
the parallel TreeSPH simulation code {\small GADGET}
\citep{Springel2001}.  It uses an entropy-conserving formulation of SPH
(Springel \& Hernquist 2002), and was supplemented with the treatment of
magnetic field in the framework of ideal magneto-hydrodynamics as
described in \citet{Dolag2002}. Since the primary purpose of the
simulation was to study the propagation of the ultra-high energy cosmic
rays in the extra-galactic magnetic fields, the simulation did not
include the effects of radiative cooling of gas, galaxy formation,
energy feedback and metal enrichment. Nevertheless the simulation data
provide useful mock samples for targeted observations of WHIM; for
example, according to numerical simulations presented in
\citet{Dave2001}, the effect of gas cooling can change the clumpiness of
WHIM. The difference between the results with and without radiative
cooling, however, is within a factor of two. Therefore, it may change the
results at some places in a quantitative manner, but as far as the
detectability and the overall distribution of WHIM (smoothed over the
angular resolution of {\it DIOS}, for instance) are concerned, such
details are not supposed to be essential. On the other hand, the
detectability of the WHIM via oxygen emission lines is sensitively
dependent on the assumed metallicity model. So we adopt an empirical
model (eq.[\ref{eq:metallicity-model}] below), which is shown to
reproduce the observed typical metallicity of intracluster medium
\citep{Cen1999b, Aguirre2001}.

For our analysis, in order to avoid a possible unrealistic effect due to
the numerical boundary, we restrict ourselves to a region within
$75h^{-1}$Mpc from our Galaxy.  Figure~\ref{fig:column_X} shows the
Hammer--Aitoff map of baryonic matter distribution in the supergalactic
coordinate. As already shown in \citet{Mathis2002} and
\citet{Dolag2003}, the simulated distribution of matter reproduces well
many known nearby structures, including the Great Attractor region, the
Great Wall structure, the Pisces--Perseus supercluster, and the Local
Void, as well as many known rich clusters such as Virgo, A3627 and Coma.
Additionally, the temperature and mass of the identified halos in the
simulation roughly agree with the observed values of their counterparts
in the real universe. Table~\ref{tab:cluster} compares the simulated and
observed values of celestial position in the supergalactic coordinate,
redshift, mass and emission weighted temperature for Coma, Virgo, A3627,
Hydra, Centaurus and Perseus clusters. The quoted masses in
Table~\ref{tab:cluster} denote the virial mass within $R_{200}$, the
radius inside which the mean density is equal to 200 times the critical
mass density for Coma, Virgo, Hydra, and Perseus, while we consider mass
within $40'$ for A3627 and $50'$ for Centaurus cluster. They are
consistent with the observed values, implying that the
simulation data are reliable enough for our present purpose. Nevertheless we
would like to emphasize that despite the very promising comparison on
large scales, many details in the simulations are not fully constrained
by the construction of the initial conditions. That is specially true
for the filamentary structure, which is certainly correct in
representing their presence in a statistical manner within the
supercluster structures, but their exact position within these complexes
cannot be compared with the real observations. Therefore predicted
emission from these structures within the superclusters can be well
compared with observations, but one should expect that the exact
location within these superclusters will not be reflected in real
observations.

Within the supergalactic coordinates, we have many galaxy clusters and
superclusters near the supergalactic equatorial plane. The upper left
and right panels of Figure~\ref{fig:equatorial} show the density and
temperature field of a 9$h^{-1}$Mpc thick slice parallel to the
Super-Galactic plane, extending from ${\rm SGZ}=-4.5h^{-1}$Mpc to ${\rm
SGZ}=4.5h^{-1}$Mpc.

\section{Oxygen Line Emission from WHIM in the Simulated Local Universe}

The goal of this paper is to examine the detectability of WHIM in the
local universe through its oxygen line emission. Naturally the expected
flux sensitively depends on the metallicity of the WHIM or more
generally of the intergalactic medium (IGM), which is very uncertain
both theoretically and observationally. In addition the simulation data
that we use do not take account of the effect of galaxy formation and
its succeeding energy feedback and metal enrichment. Thus we have to
assume a phenomenological metallicity model in subsequent analyses.  In
this paper, we adopt the following density-dependent model:
\begin{equation}
\label{eq:metallicity-model}
   Z/Z_{\odot}= {\rm Min} \left[0.2, ~ 0.02(\rho_{\rm gas}/
		    \bar{\rho}_{\rm b})^{0.3}\right],
\end{equation}
where $Z_{\odot}$ denotes the solar abundance.  This is motivated by the
radiation pressure ejection model of \citet{Aguirre2001} where the
stellar light exerts radiation pressure on the interstellar dust grains
and expels them into the hosting galactic halos and the ambient
IGM. This is the same as Model IV in Paper I except that we put an upper
limit $0.2Z_{\odot}$ so as to avoid the unacceptably large metallicity
at cluster center \citep{Cen1999b}. In practice, however, the upper
limit does not affect the result for WHIM that is of our primary
interest.  The lower left panel of Figure~\ref{fig:equatorial} indicates
the resulting metallicity map on the Super-Galactic plane, illustrating
that the prominent filamentary structures and supercluster regions
exhibit metallicity of $\log_{10}(Z/Z_{\odot})>-1.5$.

The proper identification of the oxygen emission lines from WHIM is
technically very challenging with the existing X-ray missions, and
requires a small but dedicated soft X-ray mission (Paper I).  Throughout
the paper, we assume the currently proposed specification of {\it DIOS}
\citep{Ohashi2004} which is summarized in Table
\ref{tab:mission_summary}; high energy resolution of $\Delta E=2\, {\rm
eV}$ around $E\simeq 600{\rm eV}$, and a nested 4-stage X-ray telescope
with an effective area $S_{\rm eff}=100 \, {\rm cm}^2$ and field-of-view
$\Omega_{\rm FOV}=1\, {\rm deg}^2$ ($16\times 16$ pixels in total).
Table~\ref{tab:limit} shows the detection limit of the line emission at
photon energy $E=600$ eV for {\it DIOS} in the cases of signal-to-noise
ratio $S/N=10$ and $5$ assuming three different exposure times, $T_{\rm
exp}=10^4$, $10^5$, and $10^6$ sec. Those detection limits are computed
from equation (4) of Paper I assuming that the flux of the cosmic X-ray
background (CXB) is $3\times10^{-8}$ erg s$^{-1}$ cm$^{-2}$ sr$^{-1}$
keV$^{-1}$ between 0.5 and 0.7 keV.  In what follows, we adopt the
nominal detection limit of $1\times10^{-11} {\rm
erg~s^{-1}~cm^{-2}~sr^{-1}}$ both for O{\sc vii} and O{\sc viii} lines.

As in the Paper I we calculate the O{\sc vii} and O{\sc viii} line
emission assuming that WHIM is under the collisional ionization
equilibrium. In calculating surface brightness of the line emission, we
divide the whole simulation volume into cells, each of which has
colatitudal and azimuthal extension of $\Delta {\rm SGL}=0.5^\circ$ and
$\Delta {\rm SGB}=0.5^\circ$ on the celestial plane, respectively, and
has an equally spaced redshift interval (75 bins from $z=0$ to
$z=0.03$). The line surface brightness of O{\sc vii} and O{\sc viii} for
each cell is given by
\begin{equation}
\label{eq:surface_brightness}
 S = \frac{1}{4\pi}\left(\frac{X}{m_{\rm p}}\right)^2\sum_i  
 \frac{\rho_i m_i}{(1+z_i)^4\,\Delta A_i} \,
[f_{\rm e}(T_i)]^2 \, \epsilon(T_i,Z_i),
\end{equation}
\begin{equation}
 \Delta A_i \equiv [d_{\rm A}(z_i)]^2 \, \Delta \omega ,
\end{equation}
where the summation is carried out over all gas particles (labelled $i$)
within the cell.  In the above expression, $X$ is the hydrogen mass
fraction (we adopt $0.755$), $m_{\rm p}$ is the proton mass,
$\epsilon(T,Z)$ is the oxygen line emissivities for gas temperature $T$
and metallicity $Z$ in units of the power input normalized to the
electron density as defined by \citet{Mewe1985}, $f_{\rm e}(T)$ is the
electron number density fraction relative to the hydrogen, $d_{\rm
A}(z)$ is the angular diameter distance to redshift $z$, $\Delta \omega$
is the solid angle of each cell, and $T_i$, $z_i$, $\rho_{i}$, and $m_i$
denote temperature, redshift, mass density and mass of the $i$-th gas
particle, respectively.

We note that since we assume that WHIM is in the collisional ionization
equilibrium and not affected by photoionization by CXB and UV background
radiation, both the electron number density fraction $f_{\rm e}$ and the
oxygen line emissivities $\epsilon$ are independent of gas density.
According to \citet{Kang2004}, more than 95\% of the O{\sc vii} and
O{\sc viii} line emissions come from WHIM with temperature $T>10^6$K
even under the presence of photoionization background. Actually, under
the ionization equilibrium, the emissivities $\epsilon$ of O{\sc vii}
and O{\sc viii} at $T>10^6$K is little affected by the photo-ionizing
background. Thus, as long as ionization equilibrium is assumed, the
results presented in this paper are not significantly altered even if the
effect of the photo-ionization background is considered. For WHIM with
lower temperature and lower density, however, we have to take account of
the photo-ionization effect as well as non-equilibrium treatment which we
hope to report separately.

Figure~\ref{fig:oxygen_emission} shows the simulated all-sky maps of
surface brightness for O{\sc vii} (574 eV; {\it top}) and O{\sc viii}
(653 eV; {\it middle}) line emissions in comparison with the
conventional X-ray map (0.5--2.0 keV; {\it bottom}).  At a photon energy
of $E=600$eV, the dominant interstellar absorption is due to the
Galactic oxygen as well as other metals.  The oxygen column density is
usually assumed to be in proportion to the Galactic neutral hydrogen
column density. So one can use the observed column density of neutral
hydrogen $N_{\rm HI}$ in inferring the degree of the absorption. We
consider the effect by combining the Galactic H{\sc i} map by
\citet{Dickey1990} and the absorption cross section by
\citet{Morrison1983} where the solar metal abundance is assumed. In
contrast to the 0.5--2.0 keV X-ray emission, O{\sc vii} and O{\sc viii}
emissions preferentially come from small objects like galaxy groups and
the outskirts of rich galaxy clusters and superclusters.  Furthermore,
as is clear for simulated Virgo, O{\sc vii} (574 eV) emission avoids the
very central regions of rich galaxy clusters. This is because the
emissivity of O{\sc vii} emission lines sharply drops beyond $T>10^7$ K
(see Fig. 3 of Paper I).  Figure~\ref{fig:oxygen_emission_SGplane}
depicts the O{\sc vii} and O{\sc viii} surface brightness distribution
on the supergalactic equatorial plane ($-15^\circ<{\rm SGB}<15^\circ$)
in redshift space (in plotting these maps, we use the velocity field of
the simulation data, which is smoothed over 30 neighbor particles in an
SPH manner).  A vacant strip extending at ${\rm SGL}=0^\circ$ and ${\rm
SGL}=180^\circ$ is due to the Galactic extinction.

These plots help identify several filamentary structures which have
detectable levels of oxygen emissions with {\it DIOS}; $z=0.01-0.015$
and ${\rm SGL}=190^\circ-200^\circ$ in front of A3627, $z=0.01-0.02$ and
${\rm SGL}=280^\circ-300^\circ$ in the westward of Pisces--Perseus
supercluster. In addition, we note an oxygen line emitting clump in
front of Coma at redshift $z\simeq0.01$. We will discuss the latter in
much detail so as to consider the relevance to the observational report
by \citet{Finoguenov2003} in section~\ref{sec:clump}.

If both O{\sc vii} and O{\sc viii} emissions are simultaneously
observed, one can use their line ratio as a diagnostics for the nature
of the observed WHIM, in particular its temperature.
Figure~\ref{fig:intensity_ratio} shows the contour of correlation
between the 0.5--2.0 keV band emission weighted gas temperature and
the ratio of O{\sc vii} (574 eV) and O{\sc viii} (653 eV) surface
brightness for all the regions where both O{\sc vii} and O{\sc viii}
emissions exceed the detection limits (we consider both $10^{-11}$ and
$10^{-12} {\rm erg~s^{-1}~cm^{-2}~sr^{-1}}$ for definiteness).  The
surface brightness corresponds to an aperture size of $\Delta {\rm
SGL}=\Delta {\rm SGB}=0.5^{\circ}$ ({\it left panel}) and $0.25^{\circ}$
({\it right panel}), and is projected over the redshift interval of
$\Delta z=0.0033$, which corresponds to spectral energy resolution of
DIOS ($\Delta E=2$ eV) at photon energy $E=600$ eV.

Figure~\ref{fig:intensity_ratio} implies that the temperature range of
WHIM oxygens detectable by {\it DIOS} is mainly $T=10^6-10^7$ K as
already concluded in Paper I.  The solid line indicates the relation for
the collisional ionization equilibrium. Since this is the assumption
that we adopted, the simulation data should exactly follow the relation
{\it if the smoothed cell size is sufficiently small}. In reality,
however, the effective observational resolution is determined by the
telescope angular resolution and the spectroscopic energy
resolution. Thus the inhomogeneous density and temperature structure
within the cell results in the departure from the assumed relation both
statistically and systematically. The systematic deviation from the
assumed relation seems prominent for lower temperature cells ($T<
2\times 10^6$ K).  This is mainly because the oxygen line emissions in
those cells are dominated by bright line emission regions (usually high
density and high temperature regions) within each cell while the average
temperature of each cell is lower. This interpretation is consistent
with the fact that the systematic deviation becomes weaker for analysis
using the smaller aperture size ({\it right panel}).

Figure~\ref{fig:probed_fraction} plots the fraction of baryon mass
contained in those cells where O{\sc vii} ({\it left}) and O{\sc viii}
({\it right}) line emissions exceed the given detection limit. Three
filled patterns indicate the different ranges of the temperature
mass-weighted over the ``observed'' regions. For the same reason as
described above, especially at lower temperature regions, the line
emission is not necessarily responsible for all the mass inside the
cell. Therefore, the mass fraction shown here does not follow the
detected mass precisely and it should be rather interpreted as the upper
limit of the ``detected'' fraction. These plots again indicate that more
than 90 \% of the ``detected'' mass has temperature higher than $10^6$ K
and that $\sim 60$ \% has temperature with $10^6 {\rm K} < T < 10^7 {\rm
K}$. Since baryons with $T>10^7$ K are supposed to be already detected
through continuum X-ray emission by conventional X-ray missions,
previously unexplored and detectable only through the oxygen emissions
will be about 20 \% for the nominal detection limit of {\it DIOS},
$10^{-11} {\rm erg~s^{-1}~cm^{-2}~sr^{-1}}$. It should be also
noted that the ``detected fraction'' depends on the metallicity assumed
here (eq.[\ref{eq:metallicity-model}]), and that if higher (lower)
metallicity models are assumed, we will have higher (lower) ``detected''
fraction for a given detection limit.

For the detection of WHIM through its emission, outskirts of rich galaxy
clusters are the most promising regions for the targeted survey. We
estimate the probability that O{\sc viii} (653 eV) and O{\sc vii} (574
eV) line emissions are brighter than a given limiting flux within a
$0.5^\circ\times0.5^\circ$ region around rich galaxy clusters, Coma,
Perseus, Hydra, A3627 and Centaurus clusters.  Incidentally
figure~\ref{fig:oxygen_emission} indicates that the brightest O{\sc vii}
and O{\sc viii} emitter in this simulation is Virgo cluster. However,
considering the 2eV energy resolution of XSA on-board {\it DIOS}, the
minimum redshift of an extra-galactic oxygen line emitter where it can be
discriminated from the Galactic oxygen emission is $z_{\rm min}\simeq
(2{\rm eV}/600{\rm eV})=0.0033$. Therefore, oxygen emission lines of
Virgo cluster at redshift $z=0.0038$ will be seriously contaminated by
the strong Galactic oxygen emission \citep{McCammon2002}. This is why we
do not attempt the mock observation of Virgo in the present analysis.

Figure~\ref{fig:probability} shows such probabilities for those regions
whose separation $r_{\rm s}$ from cluster centers is $r_{\rm
s}<1h^{-1}{\rm Mpc}$, $1h^{-1}{\rm Mpc}<r_{\rm s}<2h^{-1}{\rm Mpc}$, and
$2h^{-1}{\rm Mpc}<r_{\rm s}<4h^{-1}{\rm Mpc}$. Near the central regions
of galaxy clusters, O{\sc vii} emission cannot be detected so frequently
as O{\sc viii} emission because the emissivity of O{\sc vii} rapidly
drops at high temperature $T>10^7$K.  Approximately (20--30)\% area of
the outskirts of known galaxy clusters ($1h^{-1}{\rm Mpc}<r_{\rm
s}<4h^{-1}{\rm Mpc}$) in the local universe (Coma, Hydra, Centaurus,
A3627, and Perseus) exceeds the nominal detection limit of {\it DIOS}
($10^{-11} {\rm erg~s^{-1}~cm^{-2}~sr^{-1}}$) for O{\sc vii} and O{\sc
viii} emissions.

\section{Examples of Mock Spectra with {\it DIOS}}

Let us provide more specific examples of mock observation of the local
universe with {\it DIOS} including the Coma region (Figs.~\ref{fig:Coma}
and \ref{fig:Coma_spec}), the filament in front of A3627
(Figs.~\ref{fig:A3627} and \ref{fig:A3627_spec}), and a clump roughly
along the line of sight toward (simulated) Coma
(Figs.~\ref{fig:Coma_filament} and \ref{fig:Coma_filament_spec}). They
are intended to represent rich clusters, a filament extending from
clusters, and a relatively small clump of WHIM, respectively.

\subsection{Simulated Coma}

Figure~\ref{fig:Coma} depicts four maps toward the simulated Coma
cluster at redshift $z=0.023$; X-ray flux in 0.5--2 keV band,
temperature weighted by emission in the 0.5--2.0 keV band, and O{\sc
vii} (574 eV) and O{\sc viii} (653 eV) line intensities.  The
corresponding mock {\it DIOS} spectra for the selected 3 regions
(labeled 1, 2, and 3 in Fig~\ref{fig:Coma}) are shown in
Figure~\ref{fig:Coma_spec} ($500 {\rm eV}<E< 700 {\rm eV}$).  The
field-of-view of those regions is chosen as $1^\circ \times 1^\circ$.

As in Paper I, we consider the CXB and the Galactic line emission as
contaminating sources, and modeled according to \citet{Kushino2002} and
\citet{McCammon2002}, respectively. Figure~\ref{fig:Coma_spec}
represents the residual spectra after the CXB and the Galactic line
emissions are statistically subtracted. The flux intensity at an energy
range from $E$ to $E+\Delta E$ is calculated as a superposition of
spectra for SPH particles:
\begin{eqnarray}
 \label{eq:sigmapectrum}
 F(E,E+\Delta E) &=& 
\sum_i \frac{\rho_im_i}{4\pi(1+z_i)^4\Delta A_i}
\left(\frac{X}{m_{\rm p}}\right)^2 f_{{\rm e},i}^2 \cr
&\times& \int_{E(1+z_i)}^{(E+\Delta
E)(1+z_i)}P(E^\prime,T_i,Z_i)dE^{\prime},
\end{eqnarray}
where the summation performed over all SPH particles in the selected
angular region. The template spectrum, $P(E,T,Z)$, for temperature $T$
and metallicity $Z$ calculated using the SPEX software version 1.10
(http://www.sron.nl/divisions/hea/spex). 

The exposure times $T_{\rm exp}$ for each region are shown in each
panel. The region 1, the center of Coma cluster, exhibits strong
continuum and O{\sc viii} emission significantly larger than the CXB
flux, while O{\sc vii} emission is faint because of its high temperature
($T\simeq 10^8$ K). In contrast, the continuum emission from region 2,
the outskirts of Coma cluster, is much fainter than the CXB, and both of
O{\sc vii} triplet and O{\sc viii} emission lines are not
detectable. This is because the temperature in this region is still high
($T \gtsim 10^7$K) and the emission measure is too low to be
observable. Finally for region 3 which contains a small clump with lower
temperature, O{\sc viii} line emission can be detected with sufficient
signal-to-noise ratio if $T_{\rm exp}=10^5$sec, while the detection of
O{\sc vii} triplet lines requires longer exposure time.  Therefore it is
not easy to identify the oxygen line emissions from outskirts of rich
clusters, like Coma, with {\it DIOS} primarily because the temperature
is too high for the appropriate oxygen ionization states.

\subsection{the filament in front of simulated A3627}

Consider next the filament in front of simulated A3627 as an example of
prominent filamentary structure with moderate temperature.
Figure~\ref{fig:A3627} shows the maps similar to Figure~\ref{fig:Coma}
but for the sky area toward the east-side of simulated A3627
cluster. Note that A3627 is located at ${\rm SGL}=190^\circ$ and ${\rm
SGB}=7^\circ$, outside of the plots.  The filamentary structure in this
plot is located in front of A3627 as is shown in
Figure~\ref{fig:equatorial}.  Figure~\ref{fig:A3627_spec} plots mock
spectra for 3 regions marked in Figure~\ref{fig:A3627}.  For those three
regions, the continuum spectrum of the CXB contribution is significantly
larger than that of the WHIM. Nevertheless O{\sc vii} and O{\sc viii}
emission lines from this filament are very prominent in the subtracted
spectra. For region 1, O{\sc viii} emission shows up at photon energy
$E=641$ eV corresponding to redshift $z=0.018$, as well as O{\sc vii}
triplet lines at the same redshift. In the spectra of the regions 2 and
3, O{\sc viii} and O{\sc vii} triplet lines are seen at $z=0.012$. All
these line emissions come from substructures or their outskirts in the
filament. Contrary to rich clusters, the gas temperature in this
filamentary structure $10^6{\rm K}<T<10^7{\rm K}$ is indeed suitable for
identifying O{\sc vii} and O{\sc viii} lines. In other words, small
galaxy groups and/or outskirts of moderate clusters, which are expected
to have gas temperature $10^6{\rm K}<T<10^7{\rm K}$, are potentially the
most important targets in searching for WHIM signatures with {\it DIOS},
which will unveil the large-scale filamentary structure which cannot be
identified with conventional X-ray observations.

\subsection{a clump along the line of sight toward Coma region}

In plotting Figure~\ref{fig:equatorial}, we noticed the presence of a
small clump of oxygen emitting WHIM along the line of sight toward Coma
region ($z \approx 0.01$). Figure~\ref{fig:Coma_filament} is the
projection maps on the sky toward the small clump.  The center of the
clump is located at ${\rm SGL}=89^\circ$ and ${\rm SLB}=-3^\circ$,
whereas the simulated Coma (Fig.~\ref{fig:Coma}) is at ${\rm
SGL}=90^\circ$ and ${\rm SGB=7^\circ}$. Thus the physical separation
between the centers of the clump and the line-of-sight toward Coma
corresponds to $\sim 5h^{-1}$Mpc at $z=0.01$.  Since the initial
condition of the simulation is generated from real galaxy distribution
but smoothed over $5h^{-1}$Mpc, such a level of misalignment may be
reasonably accounted for by an inevitable uncertainty/fluctuation, and
it is indeed possible that they are located almost along the same line
of sight. As discussed in detail in the next section, this is why we are
interested in this particular clump.

Figure~\ref{fig:Coma_filament_spec} shows the mock spectra for the
selected three regions (labeled 1, 2, and 3) in
Figure~\ref{fig:Coma_filament}. For all of the three regions, we have
strong O{\sc viii} line emissions at $E=647$ eV, which correspond to
redshift $z=0.009$, and the O{\sc vii} triplet emission lines are also
located at the same redshift.  The peak amplitude of the X-ray surface
brightness (0.5 -- 2 keV band) of this clump is a factor of 100 times
lower than that of Coma cluster. On the other hand, O{\sc viii} emission
intensity is almost the same as Coma cluster, and O{\sc vii} intensity
is even larger.

\section{Excess emission and absorption features around the simulated 
 Coma Cluster \label{sec:clump}}

\citet{Finoguenov2003} reported the XMM-Newton observation of
peripherals of Coma cluster ($\sim$ 40 arcmin apart from the center of
Coma).  They detected O{\sc vii} and O{\sc viii} emission from a
possible WHIM clump with redshift estimated as $z=0.007\pm0.004\pm0.015$
(the first and second errors indicate statistical and systematic ones,
respectively) just in front of Coma cluster. Since the error of the
estimated redshift is large, one cannot rule out a possibility that the
oxygen emission is associated with Coma cluster ($z\simeq0.023$ or
$\sim70h^{-1}$Mpc in distance) or our Galaxy.  Nevertheless it is quite
interesting that our simulated local universe indeed has a possible
counterpart for the WHIM at an approximately right redshift $z\approx
0.01$.  Therefore it is tempting to carry out more quantitative
comparison between the simulated and the real universes.

Figure~\ref{fig:Coma_filament_prof} depicts the profiles of gas
overdensity ({\it top}), metallicity ({\it middle}) and mass weighted
temperature ({\it bottom}) along the line-of-sight toward the three
regions (1, 2, and 3).  These plots suggest that the simulated clump at
$\sim 32h^{-1}{\rm Mpc}$ in distance ($z\approx 0.01$) is characterized
by an overdensity of $\delta=10^2-10^3$, metallicity of
$Z=0.03-0.1Z_{\odot}$, and temperature of $kT=0.1-1.0$ keV.  These
physical parameters are also roughly consistent with those estimated by
\citet{Finoguenov2003} (see their Table~1), though the simulated
temperature is slightly higher than their estimation.

In order to see if we can reproduce their observed spectra, we
artificially align the centers of the simulated clump and the simulated
Coma cluster. Then we compute the mock spectra of the region 4 indicated
both in Figure~\ref{fig:Coma} and Figure~\ref{fig:Coma_filament}.  The
size of the region 4 is chosen as $12'\times12'$ so as to match the
observed region of \citet{Finoguenov2003}.  Then we add those spectra
assuming that both of them are located along the same line-of-sight
$\sim 40'$ apart from the center of the simulated Coma cluster.
Figure~\ref{fig:Coma_filament_spec2} plots the spectrum combining the
two regions. This should be compared with Figure ~2 of
\citet{Finoguenov2003} except the fact that we do not take account of
the detector response function.  At photon energy $E<1$ keV, the
emission from the clump exceeds that from Coma cluster and the total
spectrum exhibits 2--3 times higher flux at that energy range. This is
qualitatively consistent with the results by \citet{Finoguenov2003}.

We overlay in a long dashed curve the observed spectrum of ``Coma-11
field'' of \citet{Finoguenov2003} on the basis of the physical
parameters estimated by them.  The resulting curve is systematically
shifted toward the lower energy side because the estimated temperature
of the observed WHIM is $kT=0.24$ keV while our simulated clump has a
higher temperature $kT=0.62$ keV.  Thus we also attempt to calculate the
spectrum of region 4 in Figure~\ref{fig:Coma_filament} after
artificially replacing the gas temperature of the simulated clump (0.67
keV) by the observed value (0.24 keV). The resulting spectrum of the
clump is shown in Figure~\ref{fig:Coma_filament_spec}. Now it is
quantitatively consistent with the observed spectrum within a factor of
two. These results lend interesting, albeit indirect, support for the
interpretation that the simulated clump corresponds to what is reported
by \citet{Finoguenov2003}.

Consider next a possibility to detect the clump in front of Coma cluster
through its absorption feature. The detection of metal absorption lines
using the grating spectrometers enables to determine the redshift of the
clump more accurately. Thus it serves as a complementary and independent
verification of the presence of the WHIM clump.  \citet{Fujimoto2004},
for instance, reported the 2.7$\sigma$ detection of O{\sc viii}
absorption features aroud the outskirts of Virgo cluster.  Indeed
XMM-Newton is scheduled to observe the Seyfert I galaxy X-Comae located
$\sim 30$ arcmin from the center of Coma cluster with 300 ksec exposure
(J.P. Henry, private communication) in order to detect the possible
absorption feature due to the WHIM associated with Coma cluster and the
clump suggested by \citet{Finoguenov2003}.  As mentioned in the above,
the temperature of the clump ($0.24$ keV) reported by
\citet{Finoguenov2003} is lower than the simulated temperature (0.67
keV).  So again we try to compute the expected significance of the O{\sc
vii} and O{\sc viii} absorption lines for two cases; in the first case,
we adopt the simulated temperature as it is, and in the second case, we
assume that the gas clump has an isothermal temperature of 0.24 keV.

For the first case, the O{\sc vii} and O{\sc viii} column densities for
the three regions (1,2, and 3) amount to $(1.9-5.7) \times10^{15}$
cm$^{-2}$ and $(2.0-5.4) \times10^{15}$ cm$^{-2}$, respectively. Here,
we calculated O{\sc vii} and O{\sc viii} ionization fractions with the
CLOUDY code \citep{Ferland1998} assuming that the baryon in the clump is
in collisional ionization equilibrium. If we assume photo-ionization
equilibrium adopting sum of the UV background estimated by
\citet{Shull1999} and the X-ray background by \citet{Miyaji1998} as the
background photon field, the O{\sc vii} and O{\sc viii} column densities
for those regions are $(1.4-5.1) \times10^{15}$ cm$^{-2}$ and $(2.3-5.9)
\times10^{15}$ cm$^{-2}$, respectively. Considering the X-ray flux of
X-Comae of $(1.6-3.9)\times10^{-12}$ erg s$^{-1}$ cm$^{-2}$ at
(0.1--2.0) keV band and the simulated O{\sc vii} and O{\sc viii} column
densities described above, the significance of O{\sc vii} (574 eV)
absorption line obtained by XMM-Newton observation with exposure of
$3\times10^5$ seconds is expected to be (0.75--3.5)$\sigma$. Therefore,
we may marginally detect the absorption feature of the clump. On the
other hand, the detection of O{\sc viii} (653 eV) absorption will be not
significant (less than 2$\sigma$ even in the most optimistic case). We
simulate the XMM-Newton RGS observation of X-Comae through the gas clump
in the region 1 in the Figure~\ref{fig:Coma_filament}.  The resulting
spectrum is shown in the upper panel of Figure~\ref{fig:xmm_rgs}. Here
we set the exposure time to 300 ksec. At wavelength $\lambda=19.2$\AA\
and 21.8\AA, we can see very weak absorption features due to O{\sc viii}
(653 eV) and O{\sc vii} (567 eV) around redshift $z\simeq0.01$,
respectively. As expected from the computed column density of O{\sc vii}
and O{\sc viii}, the absorption features are too weak to be detected.

For the second case, on the other hand, the resulting column densities
are $(0.25-2.6) \times10^{16}$ cm$^{-2}$ and $(0.60-5.8) \times10^{16}$
cm$^{-2}$ for O{\sc vii} and O{\sc viii}, respectively. The expected
significance of O{\sc vii} and O{\sc viii} absorption lines are
$(0.69-17) \sigma$ and $(1-22) \sigma$, respectively. Therefore, in this
case, in contrast to the first case, we expect the firm detection of
both O{\sc vii} and O{\sc viii} absorption lines. The lower panel of
Figure~\ref{fig:xmm_rgs} shows the simulated XMM-Newton RGS spectrum for
the second case. All the model parameters except the gas temperature are
the same as those adopted in the upper panel. In this case, O{\sc vii}
(574 eV) and O{\sc viii} (653 eV) absorption lines will be clearly
detected with sufficient significance.

\section{Summary and conclusions}

We have explored the possibility to locate the Warm-Hot Intergalactic
Medium in the nearby universe with a dedicated soft X-ray mission to
search for cosmic missing baryons, {\it DIOS}.  A constrained
hydrodynamic simulation of the local universe has enabled us to
construct realistic mock spectra assuming the energy resolution of $\sim
2$eV and typical exposure times of $\sim 10^5$ sec.  We find that, while
Virgo cluster is too close to our Galaxy in redshift space, and its
oxygen emission may be severely contaminated by the Galactic one,
several filamentary structures around the other rich clusters in the
local universe are plausible and promising targets for the WHIM search
with {\it DIOS}. Thus their targeted survey plays a complementary role
to the blind survey for WHIM located at $0.0<z<0.3$ as discussed in
Paper I.  In those regions both O{\sc vii} and O{\sc viii} emissions are
likely to be identified simultaneously, and one can use their line ratio
as a diagnostics for the nature of the observed WHIM, in particular its
temperature. We also find that we can detect detect O{\sc vii} or O{\sc
viii} emission in approximately (20--30)\% area of the outskirts of
known galaxy clusters in the local universe (Coma, Hydra, Centaurus,
A3627, and Perseus).

Interestingly our simulated local universe indeed has a possible
counterpart for the WHIM in front of Coma cluster as reported by
\citet{Finoguenov2003}. While the temperature of the simulated clump is
a bit higher than the observational estimate, the simulated mock spectra
superposing the clump and the outskirts of the simulated Coma well
reproduce the observed feature if we set the gas temperature to the
observed value. Furthermore, we have explored a possibility to detect
the clump through its absorption feature in the spectrum of a bright
Seyfert I galaxy, X-Comae, behind Coma cluster, and we may detect the
O{\sc vii} absorption line due to the clump with the scheduled
XMM-Newton observation of X-Comae.

\bigskip 

We thank Yehuda Hoffman for calling our attention to their earlier
paper \citep{Kravtsov2002} addressing the oxygen emission signatures
from the constrained local universe simulations, and also for useful
discussion. Numerical computations presented in this paper were carried
out at ADAC (the Astronomical Data Analysis Center) of the National
Astronomical Observatory, Japan (project ID: mky05a).  KY and TF
acknowledge support from the Japan Society for the Promotion of
Science. This research was supported in part by the Grants-in-Aid by
Japan Society for the Promotion of Science (14102004, 14204017,
15340088, 15740157, 16340053).

\clearpage

\begin{table}[h]
 \begin{center}
  \caption{Simulated and observed properties of nearby clusters.\label{tab:cluster}}
  \begin{tabular}{lccccc}
   \hline\hline
   \multicolumn{1}{c}{Name} & SGL [deg] & SGB [deg] & redshift & Mass [$10^{14} M_{\odot}$] & Temperature [keV] \\
   \hline
   Coma (obs.)& 89.6 & 8.32 & 0.023 & 4.98\footnotemark[$*$] ($r<R_{200}$)&  8.0\footnotemark[$\dagger$] ($r<20'$)\\
   Coma (sim.)& 90.2 & 7.10 & 0.024 & 6.14($r<R_{200}$)  & 6.2 ($r<20'$) \\
   \hline
   Virgo (obs.)& 102 & $-3.25$ & 0.0038 & 2.04\footnotemark[$*$] ($r<R_{200}$) & 2.5\footnotemark[$\ddagger$] ($r<60'$)\\
   Virgo (sim.)& 107 & $-11.4$ & 0.0034 & 4.31 ($r<R_{200}$)& 3.8 ($r<60'$)\\
   \hline
   A3627 (obs.)& 188 & 7.04 & 0.016 & 4.0\footnotemark[$\S$] $(r<40')$ & 6.3\footnotemark[$\|$]
   ($r<20'$)\\
   A3627 (sim.)& 189 & 7.30 & 0.017 & 2.7 $(r<40')$ & 3.8 ($r<20'$) \\
   \hline
   Hydra (obs.)& 139 & $-37.5$ & 0.013 & 2.8\footnotemark[$*$]
   ($r<R_{200}$) & 3.3\footnotemark[$\#$] $(r<20')$ \\
   Hydra (sim.)& 142 & $-36.0$ & 0.011 & 2.6 ($r<R_{200}$) &  3.4
   ($r<20'$) \\
   \hline
   Centaurus (obs.) & 156 & $-11.4$ & 0.011 & 1.0\footnotemark[$**$] ($r<50'$) & 4.0\footnotemark[$\#$] ($r<10'$) \\
   Centaurus (sim.) & 159 & $-4.5$  & 0.013 & 2.4 ($r<50'$)  & 4.2 ($r<10'$)\\
   \hline
   Perseus (obs.) & 348 & $-14.1$ & 0.018 & 9.1\footnotemark[$*$]
   ($r<R_{200}$) & 6.7\footnotemark[$\|$] ($r<20'$)\\
   Perseus (sim.) & 348 & $-11.8$ & 0.016 & 8.74 ($r<R_{200}$) & 6.9 ($r<20'$)\\
   \hline
   \multicolumn{4}{@{}l@{}}{\hbox to 0pt{\parbox{85mm}{\footnotesize
   \par\noindent
   \footnotemark[$*$] \citet{Girardi1998}
   \par\noindent
   \footnotemark[$\dagger$]  \citet{Arnaud2001}
   \par\noindent
   \footnotemark[$\ddagger$] \citet{Shibata2001}.
   \par\noindent
   \footnotemark[$\S$] \citet{Tamura1998}
   \par\noindent
   \footnotemark[$\|$] \citet{De Grandi2002}
   \par\noindent
   \footnotemark[$\#$] \citet{Furusho2001}
   \par\noindent
   \footnotemark[$**$] \citet{Ettori2002}
   }\hss}}
  \end{tabular}
 \end{center}
\end{table}

\begin{table}[h]
\begin{center}
\caption{{\it DIOS} Mission summary  \label{tab:mission_summary}}
\begin{tabular}{ll}
\hline
Four-reflection X-ray telescope\\
\hline
FOV$\times$effective area & $S\Omega=$100 cm$^{2}$deg$^{2}$ at  0.6 keV\\
angular resolution  & $\sim 3$ arcmin \\
\hline
\hline
X-ray imaging spectrograph\\
\hline
energy range& $0.3 {\rm keV} <E< 1{\rm keV}$ \\
energy spectral resolution & 2eV \\
size of detector & $>$10mm$\times$10mm \\
number of pixels  & $\sim $16$\times$16  \\
FOV  & $\sim 1^{\circ}\times 1^{\circ}$  \\
\hline
\hline
Satellite system \\
\hline
orbit lifetime & $>1$ year\\
position control accuracy & $<0.5$ arcmin \\
weight& $<400$ kg \\
\end{tabular}
\end{center}
\end{table}

\begin{table}[h]
 \begin{center}
  \caption{Detection limit of line emission for the DIOS mission at
  photon energy $E=600$ eV.\label{tab:limit}}
  \begin{tabular}{ccc}
   \hline\hline
   \multicolumn{1}{c} {Exposure [s]} & limiting flux for $S/N=10$ &
   limiting flux for $S/N=5$ \\
   \multicolumn{1}{c} {}& erg s$^{-1}$ cm$^{-2}$ sr$^{-1}$ & erg s$^{-1}$ cm$^{-2}$ sr$^{-1}$ \\
   \hline
   $10^4$ & $3.7\times10^{-10}$ & $1.2\times10^{-10}$ \\
   $10^5$ & $6.0\times10^{-11}$ & $3.0\times10^{-11}$ \\
   $10^6$ & $1.4\times10^{-11}$ & $6.6\times10^{-12}$ \\
  \end{tabular}
 \end{center}
\end{table}

\begin{figure}[htbp]
 \begin{center}
  \rotatebox{270}{\FigureFile(60mm,100mm){column_density.ps}}
 \end{center}
 \caption{All-sky map of hydrogen column density in the simulated local
 universe (in the supergalactic coordinate). 
Counterparts of several clusters in the real universe are labeled in
 the map. \label{fig:column_X}}
\end{figure}

\clearpage

\begin{figure}[tbp]
 \begin{center}
  \vspace{8cm} 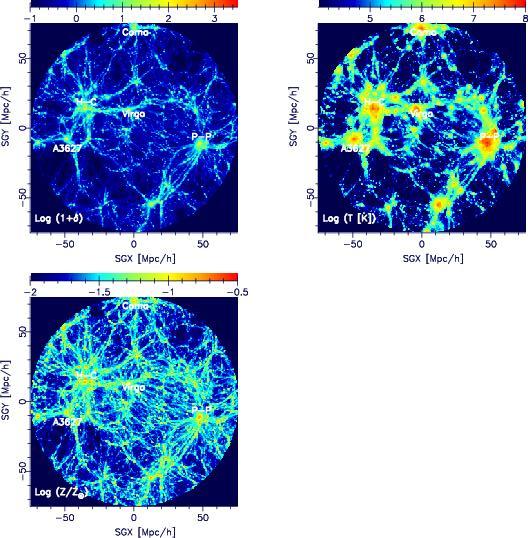
 \end{center}
\caption{Projected maps of gas density ({\it upper left}), temperature
({\it upper right}), and metallicity ({\it lower left}) in the simulated
local universe on the supergalactic plane.  The thickness of the slice
is $9h^{-1}$Mpc ($-4.5h^{-1} {\rm Mpc} < {\rm SGZ} < 4.5h^{-1} {\rm
Mpc}$). The distance is measured in real space (not in redshift space).
Labels ``H--C'' and ``P--P'' indicate the Hydra-Centaurus supercluster
and the Pisces-Perseus supercluster,
respectively.\label{fig:equatorial}}
\end{figure}

\begin{figure}[tbp]
 \begin{center}
  \vspace{8cm} 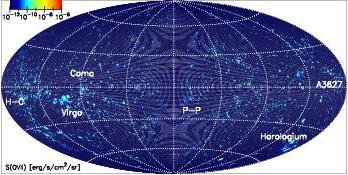

  \vspace{8cm} 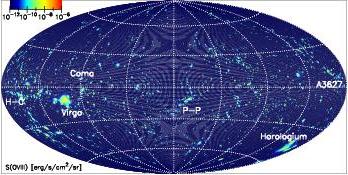

  \vspace{8cm} 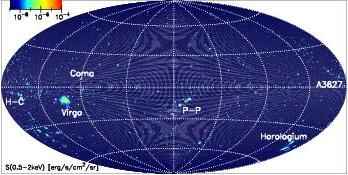
 \end{center}
 \caption{All-sky maps of O{\sc vii} (574 eV) ({\it top}), O{\sc viii}
(653 eV) ({\it middle}), and X-ray continuum (0.5--2keV) ({\it bottom})
emissions in the supergalactic coordinate.  \label{fig:oxygen_emission}}
\end{figure}

\begin{figure}[tbp]
 \begin{center}
  \vspace{8cm} 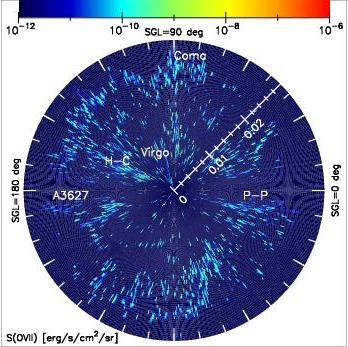

  \vspace{8cm} 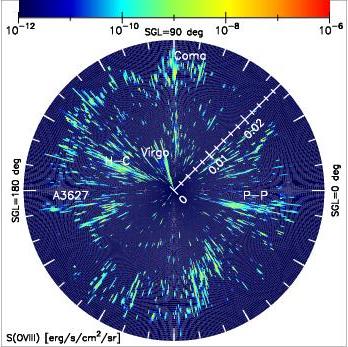
 \end{center}
 \caption{Projected surface brightness maps of O{\sc vii} ({\it upper})
 and O{\sc viii} ({\it lower}) emission lines in the simulated local
 universe on the supergalactic plane.  The thickness of the slice is
 $9h^{-1}$Mpc ($-4.5h^{-1} {\rm Mpc} < {\rm SGZ} < 4.5h^{-1} {\rm
 Mpc}$). The distance is measured in redshift
 space ($0<z<0.03$). \label{fig:oxygen_emission_SGplane}}
\end{figure}

\begin{figure}[tbp]
 \begin{center}
  \FigureFile(140mm,140mm){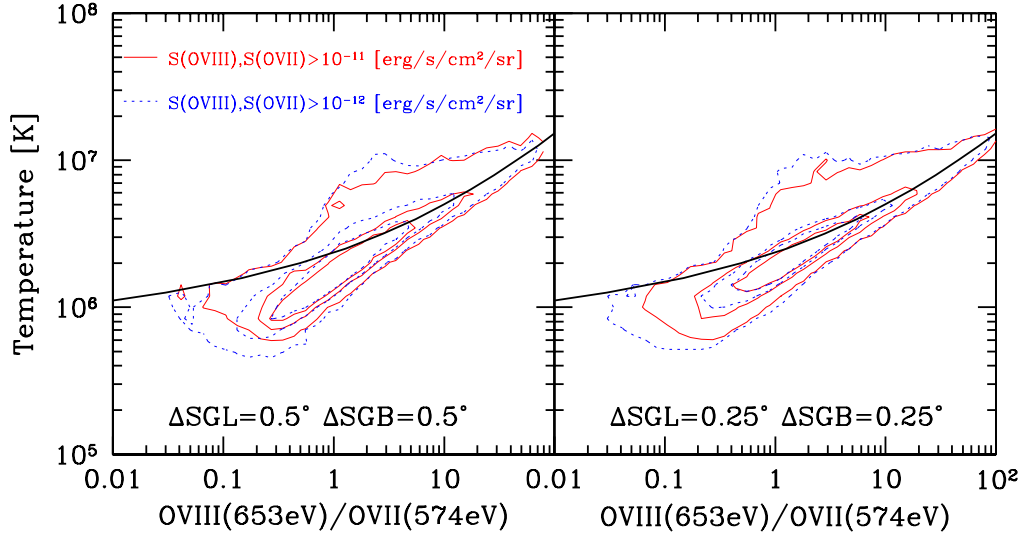}
 \end{center}
 \caption{Relation between the ratio of O{\sc vii} (574 eV) and O{\sc
 viii} (653 eV) line intensity and baryon temperature for the regions
 where both the surface brightness of O{\sc vii} (574 eV) and O{\sc
 viii} (653 eV) exceed the threshold $10^{-11}$ and $10^{-12}$ erg
 s$^{-1}$ cm$^{-2}$ sr$^{-1}$ simultaneously. The three contours from
 inside to outside indicate that the 30, 60, 90\% of the all regions are
 enclosed inside them. The size of observed area is set to $\Delta {\rm
 SGL}=\Delta {\rm SGB}=0.5^{\circ}$ ({\it left}) and
 $0.25^{\circ}$ ({\it right}), and the redshift depth is set to
 $\Delta z=0.0033$ in both panels. The solid line indicates the
 theoretical relation under the assumption of collisional ionization
 equilibrium.  \label{fig:intensity_ratio}}
\end{figure}

\begin{figure}[tbp]
 \begin{center}
  \FigureFile(160mm,100mm){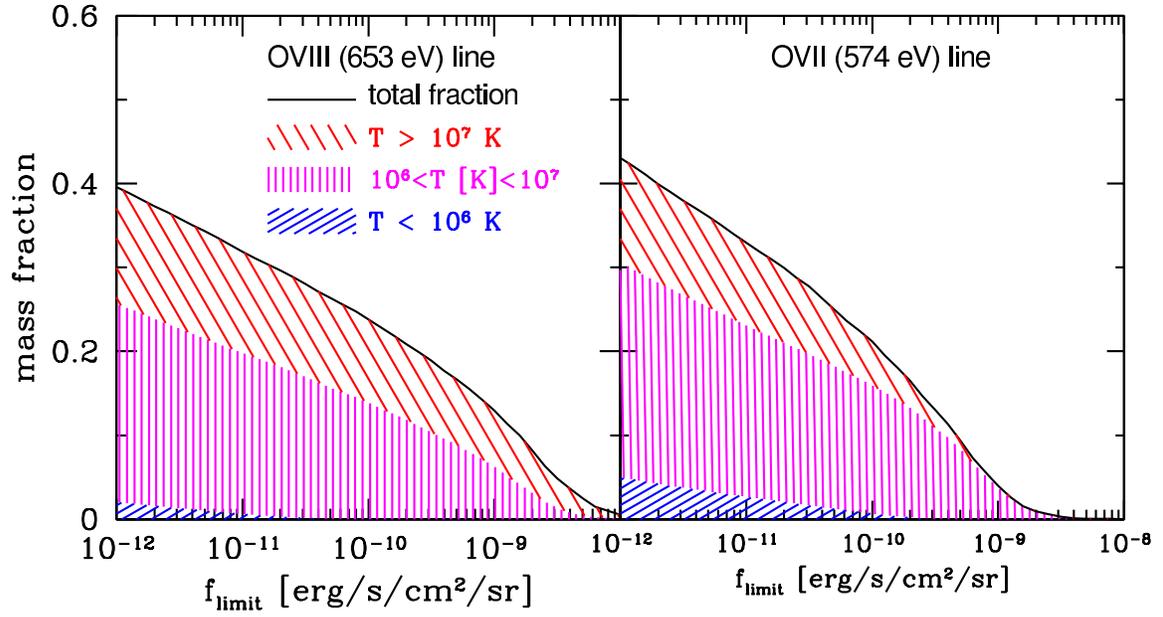}
 \end{center}
 \caption{Baryon mass fraction detected through O{\sc vii} (574 eV)
 ({\it right}) and O{\sc viii} (653 eV) ({\it left}) emission
 lines. Contributions of baryons with $T<10^6$ K, $10^6 {\rm K}
 <T<10^7 {\rm K}$, and $T>10^7 {\rm K}$ are shown
 separately.\label{fig:probed_fraction}}
\end{figure}

\begin{figure}[tbp]
 \begin{center}
  \FigureFile(160mm,100mm){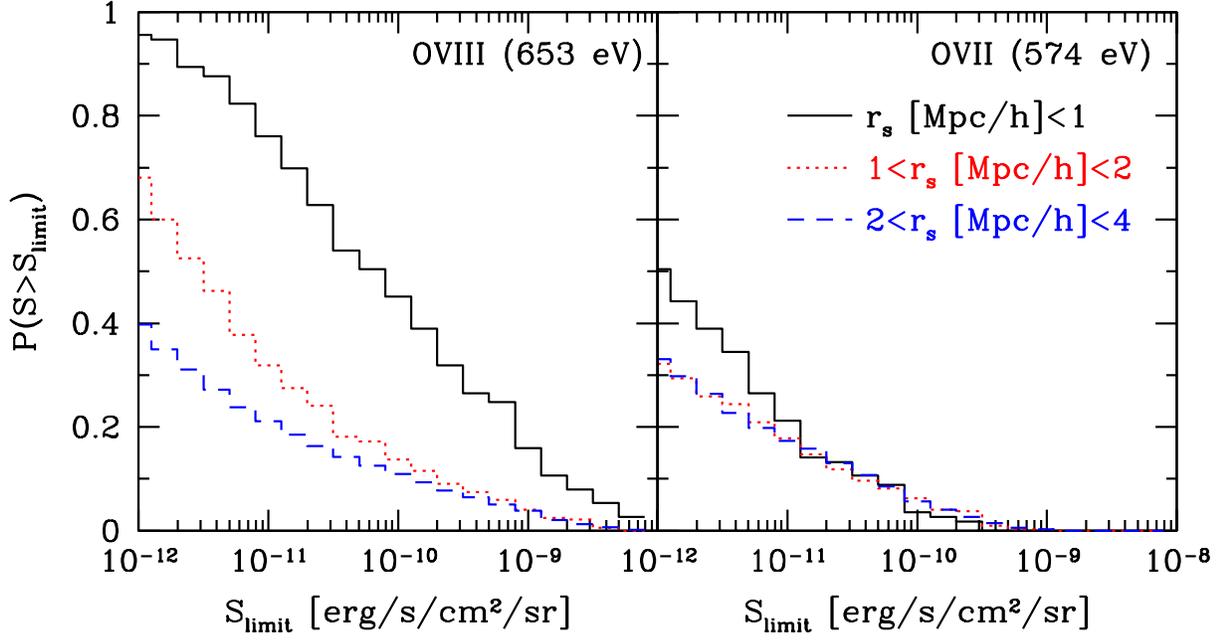}
 \end{center}
 \caption{Cumulative distribution of the oxygen flux for a
 $0.5^\circ\times0.5^\circ$ region separated from rich galaxy clusters;
 ({\it Left}) for O{\sc viii}, and ({\it Right}) for O{\sc vii}.
 Different lines correspond to the different regions of radius $r_s$
 away from the center of those clusters; $r_s < 1\, h^{-1}$Mpc ({\it
 solid}), $1\, h^{-1} < r_s < 2\, h^{-1}$Mpc ({\it dotted}), and $2\,
 h^{-1} < r_s < 4\, h^{-1}$Mpc ({\it dashed}).  \label{fig:probability}}
\end{figure}
\clearpage

\begin{figure}[h]
 \begin{center}
  \vspace{8cm}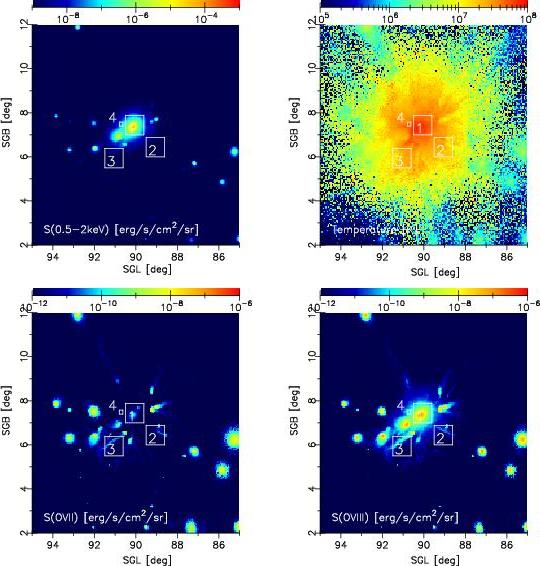
 \end{center}
 \caption{Maps of soft X-ray (0.5--2 keV) ({\it upper left}), emission
 weighted temperature ({\it upper right}), O{\sc vii} (574 eV) ({\it
 lower left}) and O{\sc viii} (653 eV) ({\it lower right}) toward
 simulated Coma cluster. \label{fig:Coma}}
 \begin{center}
  \FigureFile(100mm,100mm){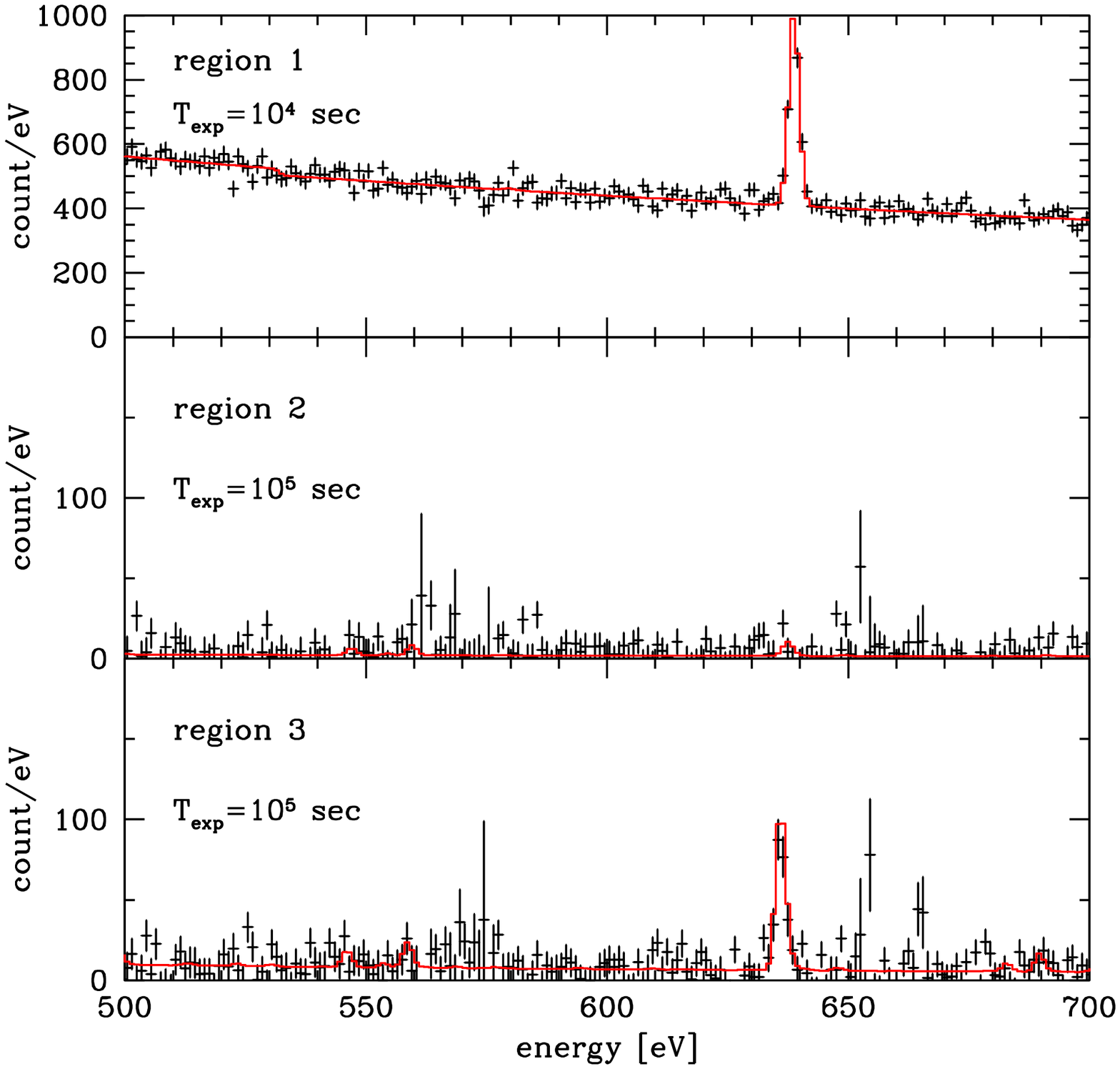}
 \end{center}
 \caption{Spectra of the 3 regions marked in Figure~\ref{fig:Coma} after
 the CXB and the Galactic emission are subtracted.\label{fig:Coma_spec}}
\end{figure}

\begin{figure}[tbp]
 \begin{center}
  \vspace{8cm}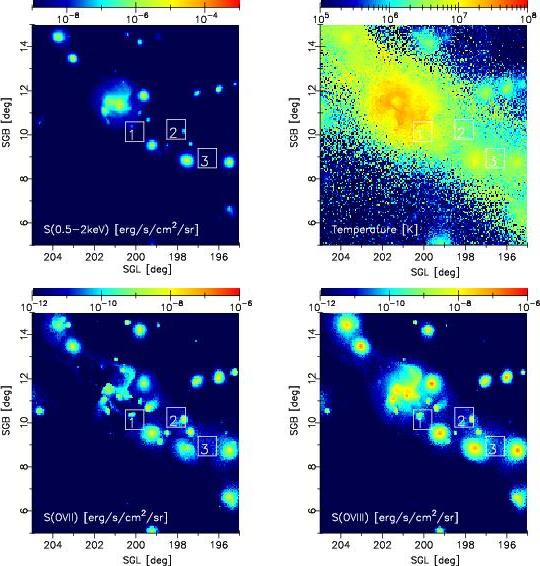
 \end{center}
 \caption{Maps of soft X-ray (0.5--2 keV) ({\it upper left}), emission
weighted temperature ({\it upper right}), O{\sc vii} (574 eV) ({\it
lower left}) and O{\sc viii} (653 eV) ({\it lower right}) toward the
simulated A3627 cluster.  \label{fig:A3627}}
 \begin{center}
  \FigureFile(100mm,100mm){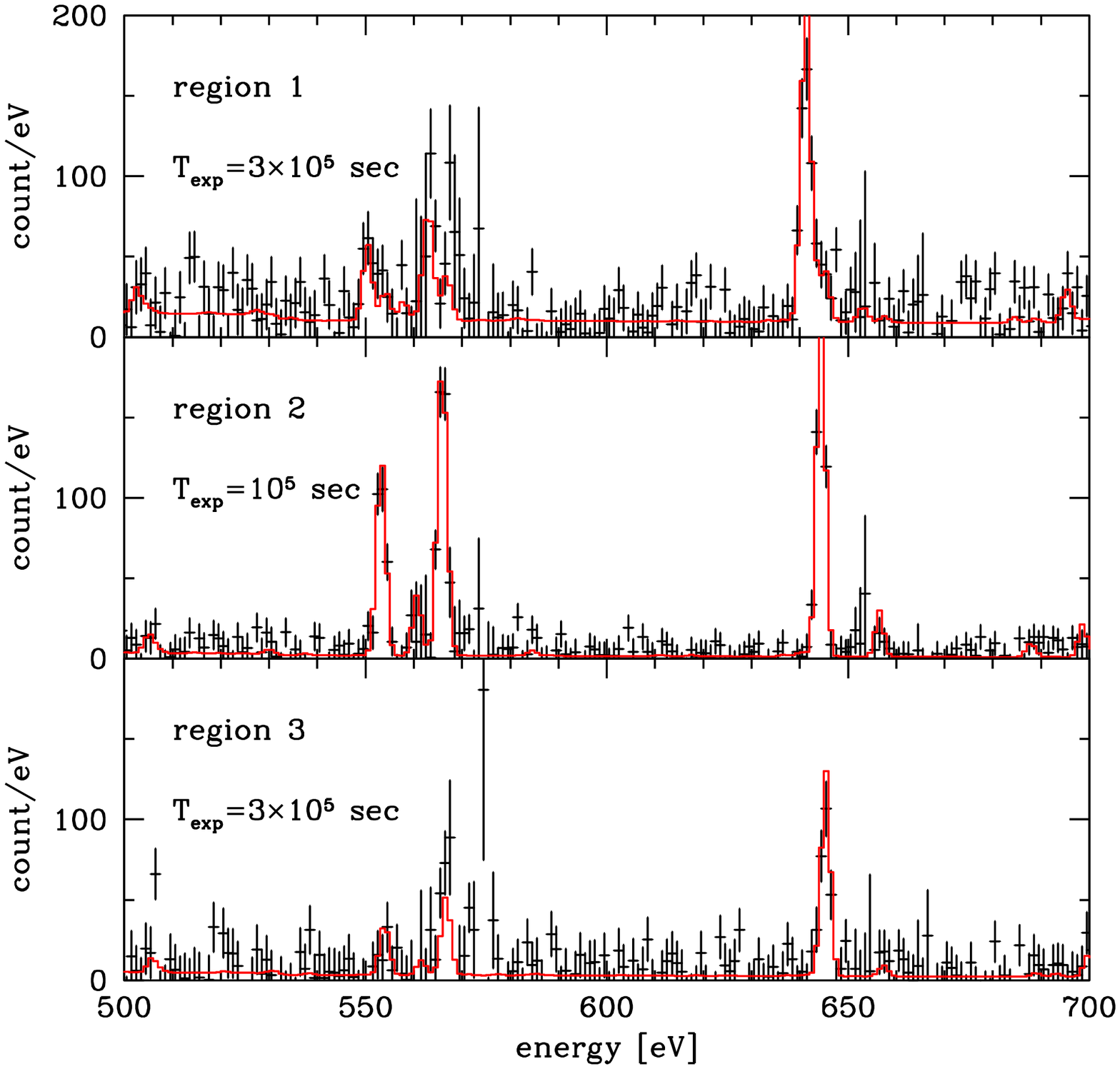}
 \end{center}
 \caption{Spectra of the 3 regions marked in Figure~\ref{fig:A3627} after
 the CXB and the Galactic emission are subtracted.\label{fig:A3627_spec}}
\end{figure}

\begin{figure}[tbp]
 \begin{center}
  \vspace{8cm} 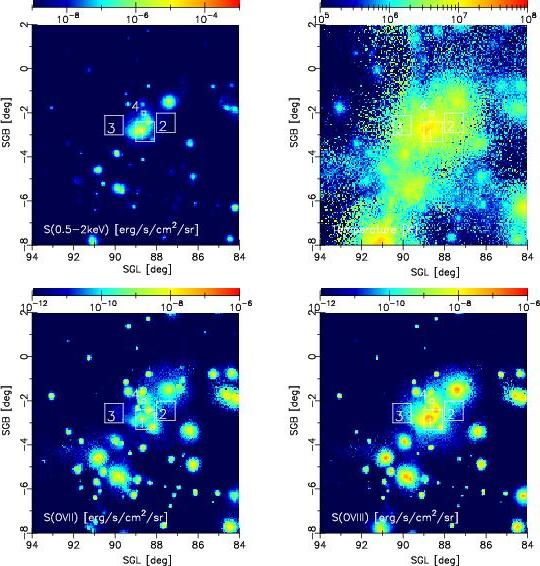
 \end{center}
 \caption{Maps of soft X-ray (0.5--2 keV) ({\it upper left}), emission
 weighted temperature ({\it upper right}), O{\sc vii} (574 eV) ({\it
 lower left}) and O{\sc viii} (653 eV) ({\it lower right}) toward
 the clump in front of
 Coma cluster.\label{fig:Coma_filament}}
 \begin{center}
  \FigureFile(100mm,100mm){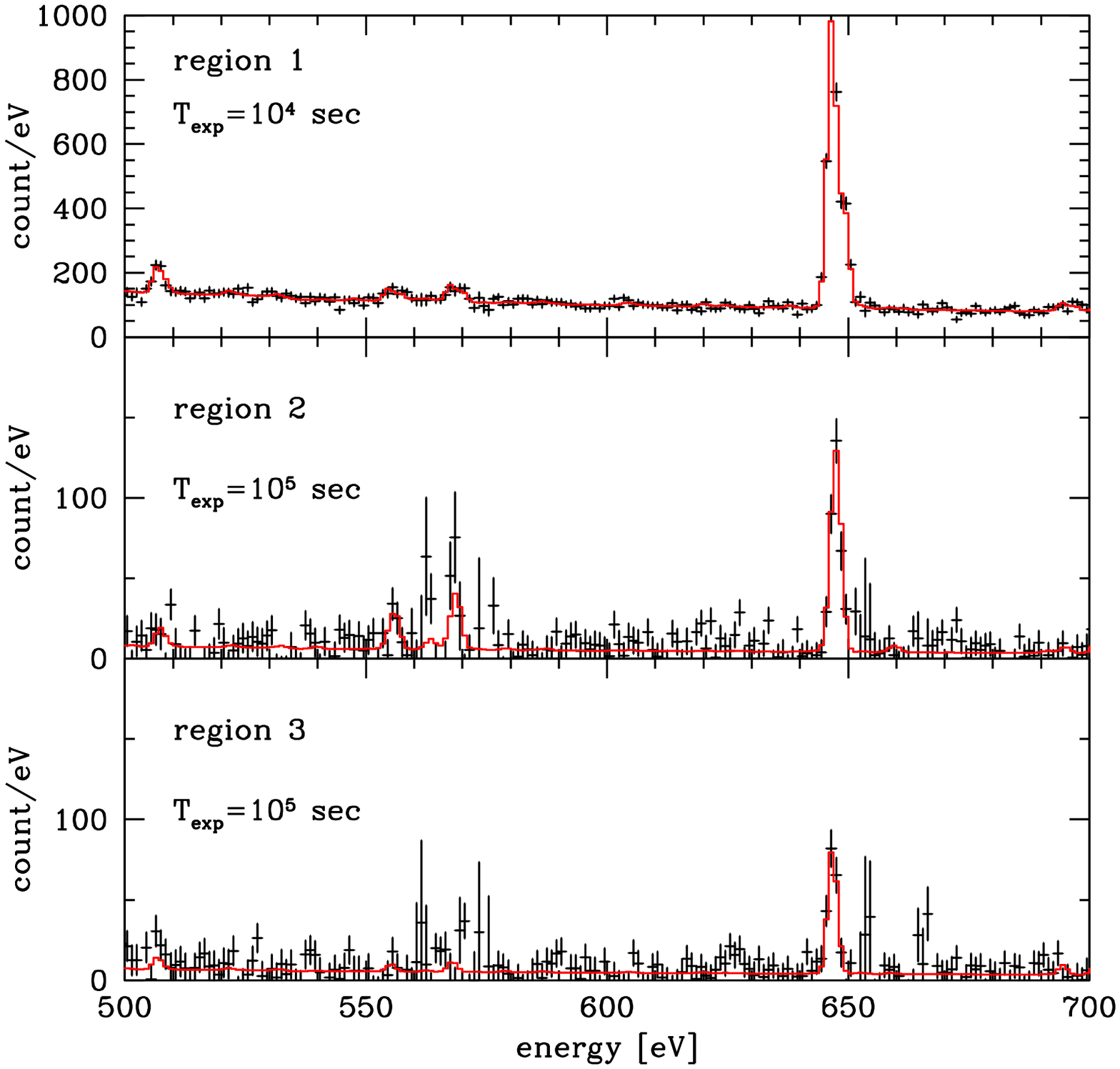}
 \end{center}
 \caption{Spectra of the 3 regions marked in
 Figure~\ref{fig:Coma_filament} after the CXB and the Galactic emission
 are subtracted.\label{fig:Coma_filament_spec}}
\end{figure}

\clearpage

\begin{figure}[tbp]
 \begin{center}
  \FigureFile(120mm,120mm){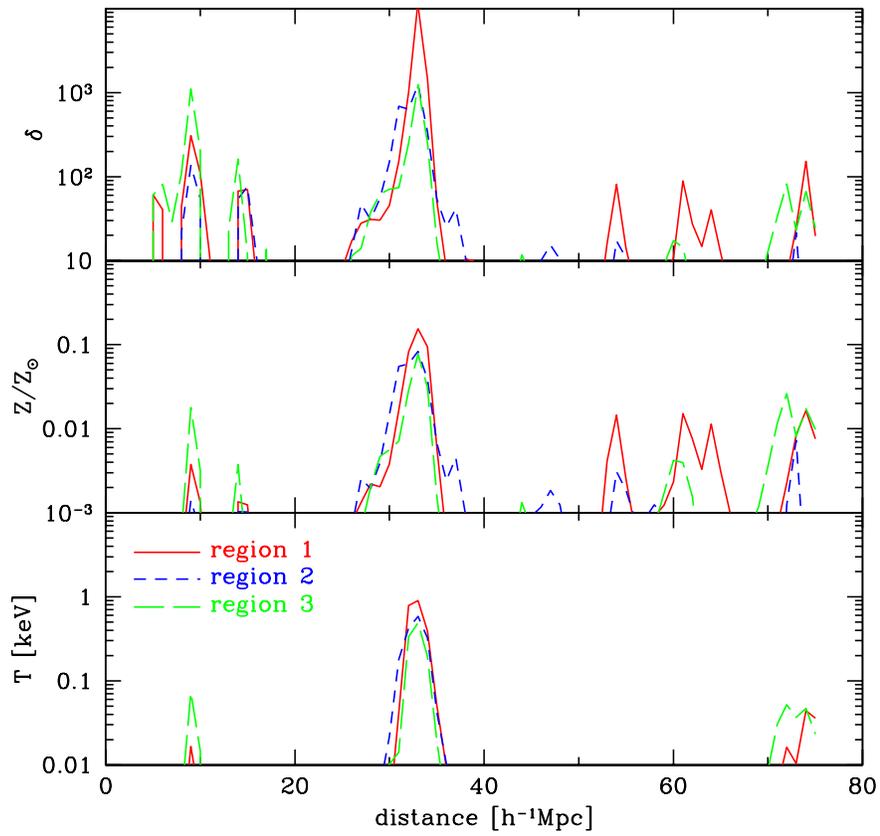}
 \end{center}
 \caption{Profiles of overdensity ({\it top}), metallicity ({\it
 middle}), and mass-weighted temperature ({\it bottom}) for the 3
 regions (labeled 1, 2, and 3) marked in Figure~\ref{fig:Coma_filament}.
 \label{fig:Coma_filament_prof}}
\end{figure}

\begin{figure}[tbp]
 \begin{center}
  \FigureFile(120mm,120mm){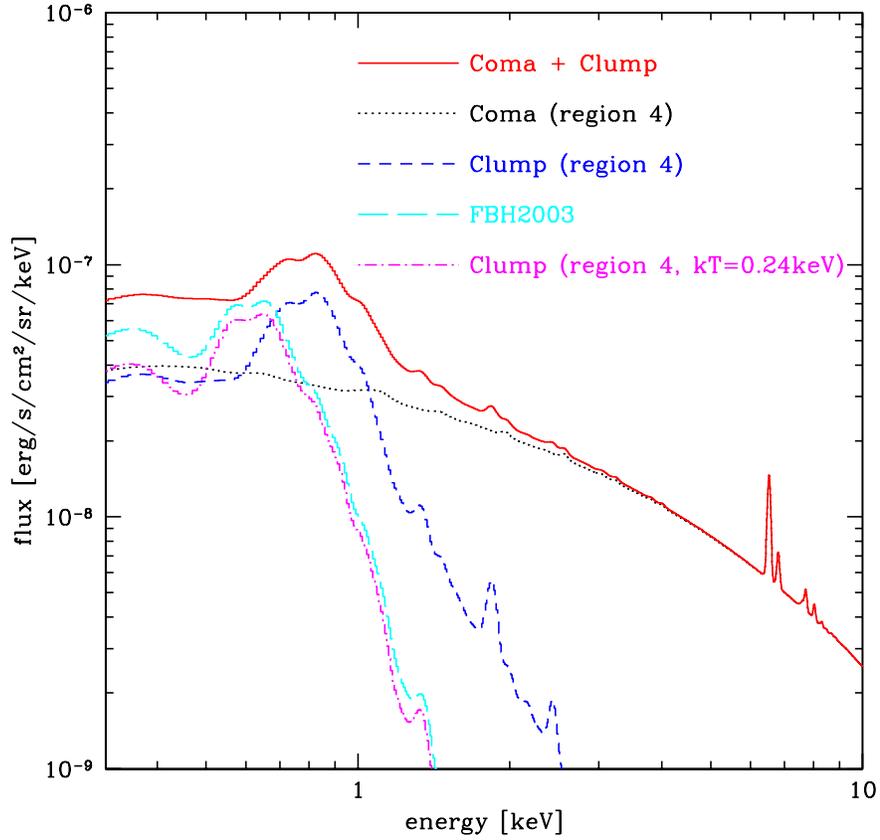}
 \end{center}
 \caption{Simulated X-ray spectra for the region 4 in the outskirts of
 simulated Coma cluster (Figure~\ref{fig:Coma}) and in the simulated
 clump in front of Coma (Figure~\ref{fig:Coma_filament}). The spectrum
 using the physical parameters of the filament reported by
 \citet{Finoguenov2003} is also shown for comparison in the long-dash
 line. The dotted-dash line indicates the spectrum for region 4 in
 Figure~\ref{fig:Coma_filament} but the baryon temperature is set to
 $kT=0.24$ keV. \label{fig:Coma_filament_spec2}}
\end{figure}

\begin{figure}[tbp]
 \begin{center}
  \FigureFile(120mm,120mm){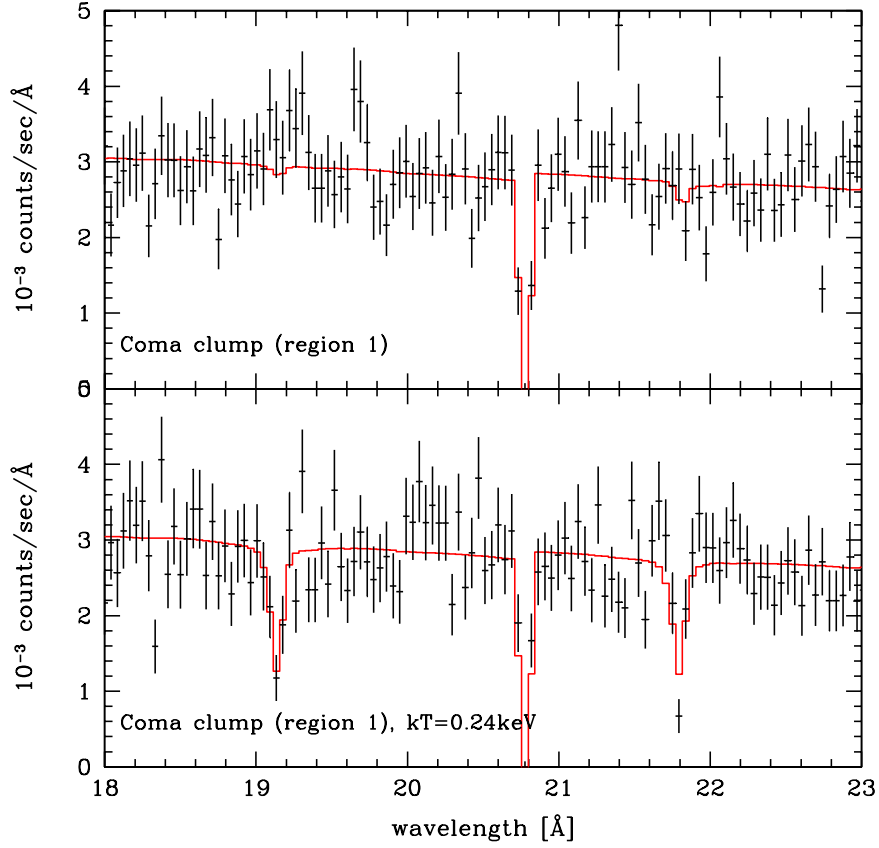}
 \end{center}
 \caption{Simulated XMM-Newton RGS spectra of absorption lines in the
 spectrum of X-Comae induced by the simulated gas clump in front of Coma
 cluster along the line-of-sight toward region 1 in
 Figure~\ref{fig:Coma_filament}. In the {\it upper} panel, the simulated
 temperature of the clump is adopted as it is, while we assume the
 temperature of 0.24 keV estimated by \citet{Finoguenov2003} in the
 lower panel. Note that the absorption-like features at 20.8\AA\ in both
 panels are not physical but simply due to the gap of CCDs on the RGS
 detector. Solid lines are the theoretically expected spectra and points
 with error-bars indicate the simulated photon
 counts. \label{fig:xmm_rgs}}
\end{figure}

\end{document}